\documentclass[showpacs,epsf,twocolumn,superscriptaddress]{revtex4-1}
\usepackage{float}
\usepackage{amsmath}
\usepackage{color,graphicx}
\usepackage{hyperref}

\begin{document}

\title{Anomalous Hall effect in half-metallic Heusler compound Co$_{2}$Ti$X$ ($X$=Si, Ge)}

\author{Shubhankar Roy}
\affiliation{Saha Institute of Nuclear Physics, HBNI, 1/AF Bidhannagar, Kolkata 700 064, India}
\affiliation{Vidyasagar Metropolitan College, 39, Sankar Ghosh Lane, Kolkata 700 006, India}
\author{Ratnadwip Singha}
\affiliation{Saha Institute of Nuclear Physics, HBNI, 1/AF Bidhannagar, Kolkata 700 064, India}
\author{Arup Ghosh}
\affiliation{Saha Institute of Nuclear Physics, HBNI, 1/AF Bidhannagar, Kolkata 700 064, India}
\author{Arnab Pariari}
\affiliation{Saha Institute of Nuclear Physics, HBNI, 1/AF Bidhannagar, Kolkata 700 064, India}
\author{Prabhat Mandal}
\affiliation{Saha Institute of Nuclear Physics, HBNI, 1/AF Bidhannagar, Kolkata 700 064, India}

\email{prabhat.mandal@saha.ac.in}

\date{\today}

\begin{abstract}
Though Weyl fermions have recently been observed in several materials with broken inversion symmetry, there are very few examples of such systems with broken time reversal symmetry. Various Co$_{2}$-based half-metallic ferromagnetic Heusler compounds are lately predicted to host Weyl type excitations in their band structure. These magnetic Heusler compounds with broken time reversal symmetry are expected to show a large momentum space Berry curvature, which introduces several exotic magneto-transport properties. In this report, we present systematic analysis of experimental results on anomalous Hall effect (AHE) in Co$_2$Ti$X$ ($X$=Si and Ge). This  study is an attempt to understand the role of Berry curvature on AHE in Co$_2$Ti$X$ family of materials. The anomalous Hall resistivity is observed to scale  quadratically with the longitudinal resistivity for both the compounds. The detailed analysis indicates that in anomalous Hall conductivity, the intrinsic Karplus-Luttinger Berry phase mechanism dominates over the extrinsic skew scattering and side-jump mechanism.
\end{abstract}
\maketitle

\section{Introduction}

Co$_2$-based Heusler compounds have enticed an immense interest in condensed matter physics due to their high Curie temperature and tunability of electronic and magnetic properties \cite{Heusler,Trudel}. Especially, the fascinating half-metallic character \cite{Groot,Galanakis}, makes these materials very promising candidates for the spintronics applications \cite{Chappert}. Recently, Co$_2$-based Heusler compounds have also been predicted to host Weyl fermions \cite{Chang,Wang}. While there are several topological Weyl semimetals with broken inversion symmetry \cite{Lv,Xu1,Xu2}, presence of such semimetallic state in magnetic systems due to broken time reversal symmetry is extremely rare \cite{Borisenko}. The Berry curvature associated with topologically non-trivial band structure, leads to novel magneto-transport properties and several exotic phenomena such as anomalous Hall effect (AHE), chiral anomaly, and anomalous Nernst effect, etc \cite{Suzuki,Singha,Hirschberger,Guin}.

Usually, in a ferromagnetic system, the measured Hall resistivity has contributions from both ordinary Hall effect and AHE. The anomalous Hall resistivity component is proportional to the spontaneous magnetization ($M$). Although the basic theory of AHE has already been reviewed extensively \cite{Nagaosa,Xiao}, the renewed interest in this phenomenon has emerged due to its strong correlation with Berry phase. It is well established that three possible mechanisms are responsible for AHE \cite{Nagaosa}. In order to explain AHE, Karplus and Luttinger were first to propose a model (intrinsic KL mechanism), which takes into account the role of spin-orbit interaction in band structure calculation of ferromagnetic metals \cite{Karplus}. According to KL mechanism, the anomalous Hall resistivity ($\rho_{xy}^{A}$) is expected to show a quadratic dependence on longitudinal resistivity ($\rho_{xx}$).  Extrinsic mechanisms such as skew-scattering and side-jump or asymmetric scattering of conduction electrons by the spin-orbit coupled impurities can also result in AHE \cite{Smit,Smit2,Berger}. While for skew-scattering, the $\rho_{xy}^{A}$ is linearly proportional to $\rho_{xx}$, side-jump scattering is proportional to $\rho_{xx}$$^2$ as in the case intrinsic KL mechanism. The intrinsic KL mechanism has in fact a direct link to the Berry-phase effects on occupied electronic Bloch states \cite{Nagaosa2,Jungwirth}.

There are several theoretical and experimental reports on the structural, electronic, and magnetic properties of Co$_2$-based Heusler compounds  \cite{Barth,Roy}. The spin-resolved band structure calculations show that the majority spin-band has a metallic character, whereas the minority spin-band exhibits semiconducting behavior with a band gap of about 0.5 eV at the Fermi level \cite{Chang,Barth}. Recent band structure calculations for these compounds show that the inclusion of the spin-orbit coupling (SOC) leads to novel topological Weyl semimetal state with time reversal symmetry breaking \cite{Chang}. However, a systematic study on this topologically non-trivial state through transport experiments is still lacking. In the present work, we report the AHE associated with Berry curvature in Heusler compounds Co$_2$Ti$X$ ($X$=Si and Ge).

\begin{figure}
\includegraphics[width=0.48\textwidth]{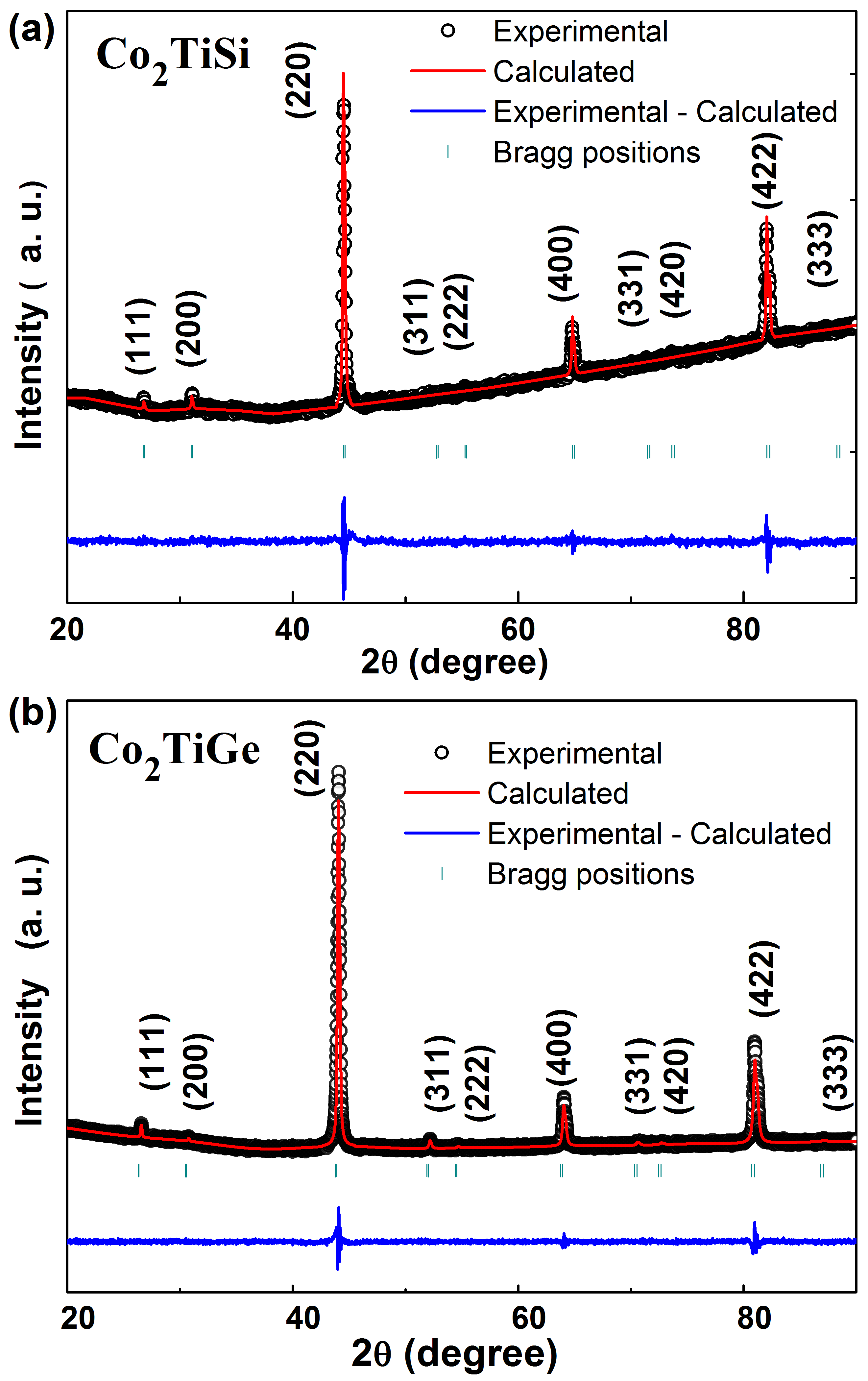}
\caption{(Color online) Rietveld profile refinement of the XRD pattern of powdered samples of (a) Co$_{2}$TiSi and (b) Co$_{2}$TiGe. Black circles($\circ$) are experimental data, red line($\textcolor{red}{-}$) is the calculated pattern, blue line($\textcolor{blue}{-}$) is the difference between experimental and calculated intensities, and vertical lines($\textcolor{cyan}{\mid}$) show the Bragg positions.}\label{rh}
\end{figure}

\section{Experimental details}

Co$_{2}$Ti$X$ ($X$=Si and Ge) compounds were prepared by arc melting the stoichiometric amounts of high purity of its constituents in a highly purified argon (Ar) atmosphere. In order to avoid any contamination with oxygen, the sample chamber was purged with high purity Ar and evacuated. This procedure was repeated several times. Furthermore, we have heated Ti pieces close to its melting point inside the vacuum chamber before arc melting the compound. To achieve chemical homogeneity, the ingot was remelted five to six times and flipped before each new melting step. The obtained ingot was then sealed in an evacuated quartz tube and annealed for 3 weeks at 950 $^{\circ}$C. Phase purity and structural analysis for both the samples were done using the powder x-ray diffraction (XRD) technique with Cu K$_{\alpha}$ radiation in a Rigaku x-ray diffractometer (TTRAX III). The magnetization measurements were performed using a superconducting quantum interference device-vibrating sample magnetometer (SQUID-VSM) (MPMS 3, Quantum Design) in magnetic fields up to 7 T. The sample used for the magnetic measurements is of approximate dimensions 0.3$\times$0.4$\times$4 mm$^{3}$. To minimize the demagnetization effect, the external magnetic field was applied along the longest sample direction. The magnetization data were recorded over the temperature range from 2 to 380 K. To achieve good thermal equilibrium, we have stabilized each temperature for 30 minutes. For each $M$($H$) isotherm, the magnetic field was increased from 0 to 7 T and then reduced to zero. We didn't observe any difference in $M(H)$ between the increasing and decreasing field. The transport measurements were performed in a 9 T physical property measurement system (Quantum Design) using the ac-transport option. For both the resistivity and Hall measurements, the electrical contacts were made in the four-probe configuration using conducting silver paste and gold wires. In order to eliminate any contribution due to the small misalignment of voltage probes, the final Hall resistivity was obtained from the difference between measured Hall resistivity for positive and negative magnetic field.

\section{Results and Discussions}

\begin{figure}
\includegraphics[width=0.48\textwidth]{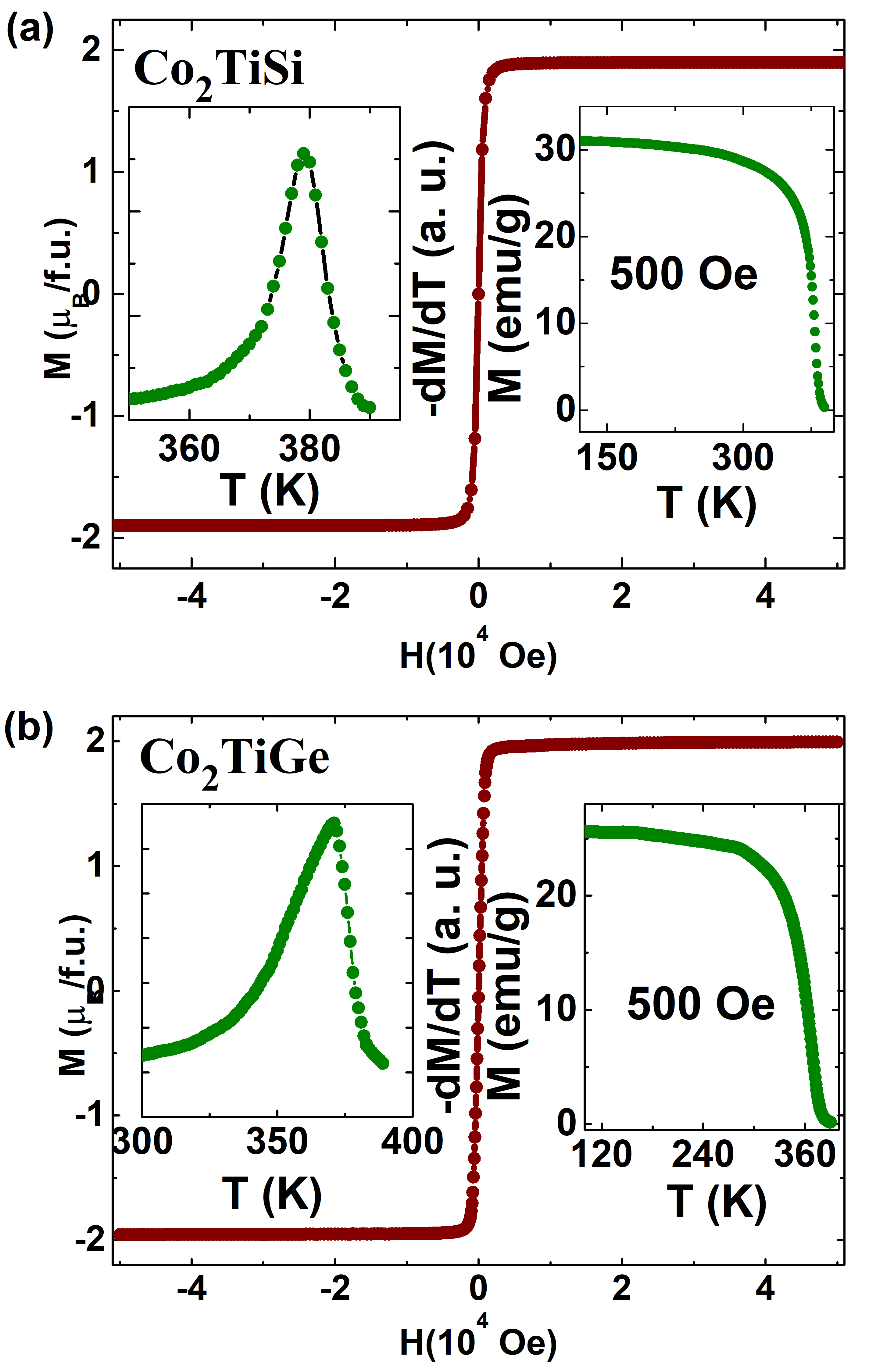}
\caption{(Color online) Magnetic field dependence of the magnetization at 2 K for (a) Co$_{2}$TiSi and (b) Co$_{2}$TiGe. Right insets: Temperature dependence of the magnetization measured at 500 Oe. Left insets: $dM$/$dT$ versus temperature.}\label{rh}
\end{figure}

Co$_2$Ti$X$  compounds crystallize in cubic $L$2$_1$ structure (space group: F\textit{m}$\bar{3}$\textit{m}), which consists of four inter-penetrating face-centered-cubic lattices along (111) direction. The crystallographic positions of Co atoms are (0, 0, 0) and (1/2, 1/2, 1/2) and those of the Ti and Ge atoms are (1/4, 1/4, 1/4) and (3/4, 3/4, 3/4), respectively. The Rietveld profile refinement of the XRD patterns (Fig. 1) confirms that the prepared materials are single phase in nature. The refined lattice parameters, $a$=$b$=$c$=5.744(1) {\AA} and $a$=$b$=$c$=5.811(1) {\AA} for Co$_{2}$TiSi and Co$_{2}$TiGe, respectively, are in good agreement with the earlier reports \cite{Barth,Barth2}.

\begin{figure}
\includegraphics[width=0.48\textwidth]{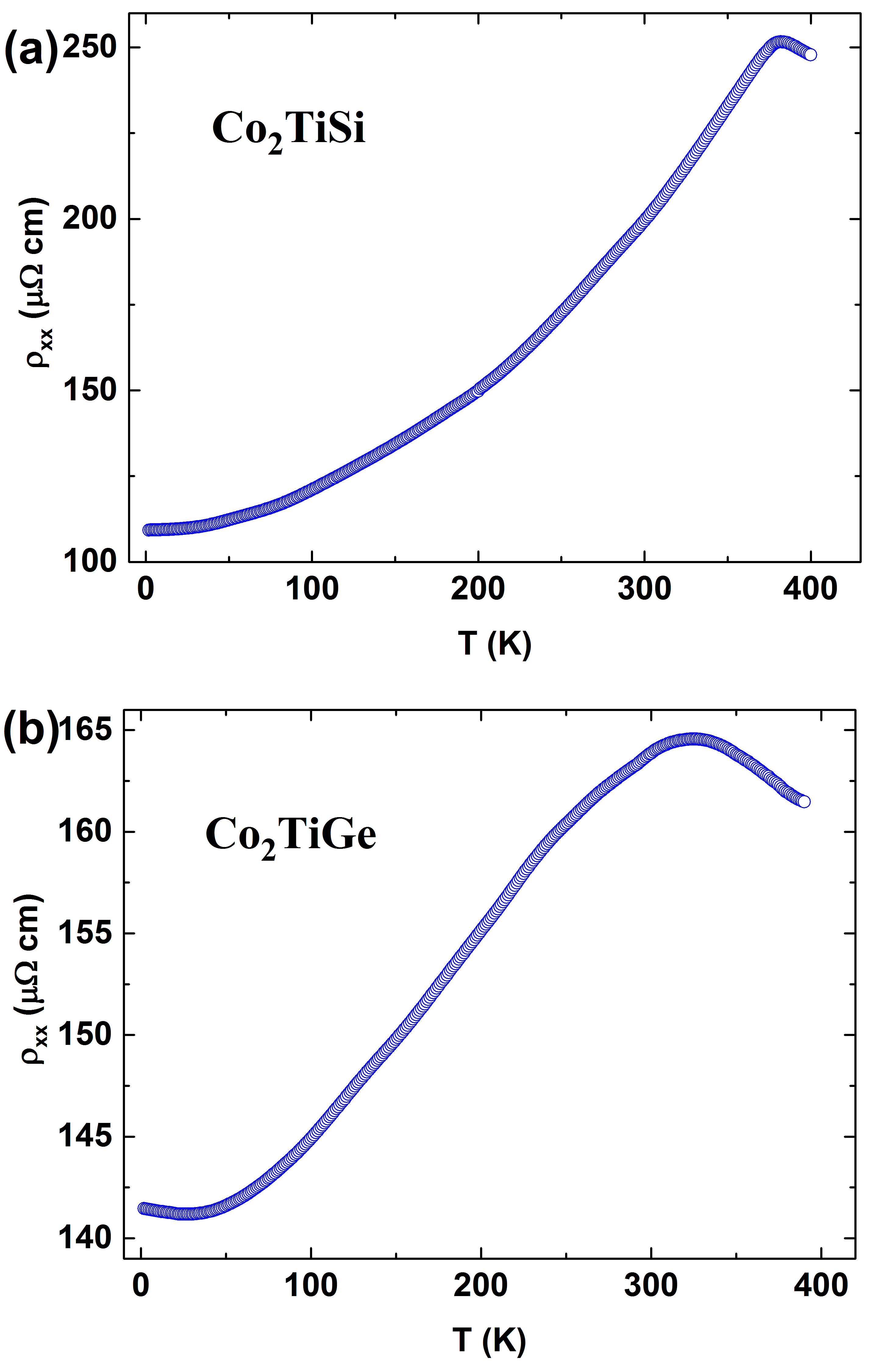}
\caption{(Color online) Temperature dependence of the longitudinal resistivity ($\rho_{xx}$) for (a) Co$_{2}$TiSi and (b) Co$_{2}$TiGe.}\label{rh}
\end{figure}

The magnetic field dependence of dc-magnetization at 2 K for Co$_{2}$TiSi and Co$_{2}$TiGe are shown in Figs. 2(a) and 2(b), respectively. The observed behavior is quite similar to the earlier reports \cite{Barth,Barth2}. The Co$_2$-based Heusler alloys, which are half-metallic ferromagnets, exhibit the Slater-Pauling-type behavior of the magnetization as given by $M_{p}$ = ($Z_{p}$ - 24) $\mu_{B}$/f.u. Here, $M_{p}$ is the total magnetic moment and $Z_{p}$ is the total number of valence electrons in the unit cell of the compound. For Co$_{2}$Ti$X$, the value of $Z_{p}$ is 26. Therefore, according to the above relation, the total magnetic moment should be 2 $\mu_{B}$/f.u. From  Figs. 2(a) and 2(b), the value of saturation magnetization ($M_S$) is estimated to be $\sim1.89$ $\mu_{B}$/f.u. for Co$_{2}$TiSi  and $\sim1.99$ $\mu_{B}$/f.u. for Co$_{2}$TiGe, which are consistent with the Slater-Pauling rule. In both figures, we have plotted the temperature dependence of $M$ in right inset, measured at 500 Oe and the corresponding temperature-derivative in the left inset. It is clear from the right insets of Figs. 2(a) and 2(b) that the magnetization curve exhibits  a continuous or second order ferromagnetic to paramagnetic phase transition. From the left insets, the Curie temperature ($T_{C}$), defined as the temperature where -d$M$/d$T$ shows a peak, is estimated to be around 380  and 371 K for Co$_{2}$TiSi and Co$_{2}$TiGe, respectively.\\

\begin{figure}
\includegraphics[width=0.48\textwidth]{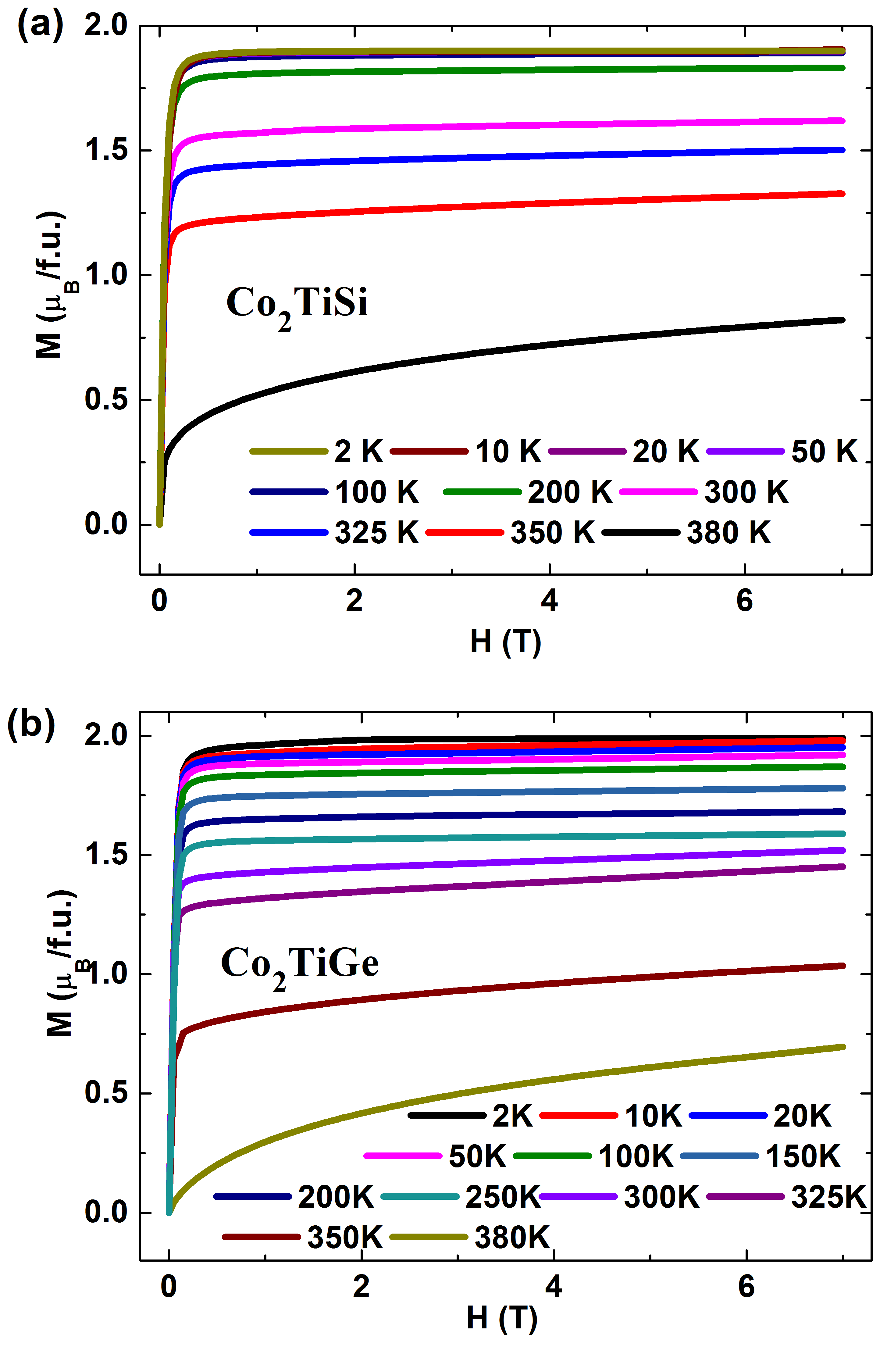}
\caption{(Color online) Magnetic field dependence of the magnetization ($M$ vs. $H$) for (a) Co$_{2}$TiSi and (b) Co$_{2}$TiGe.}\label{rh}
\end{figure}

\begin{figure*}
\includegraphics[width=0.9\textwidth]{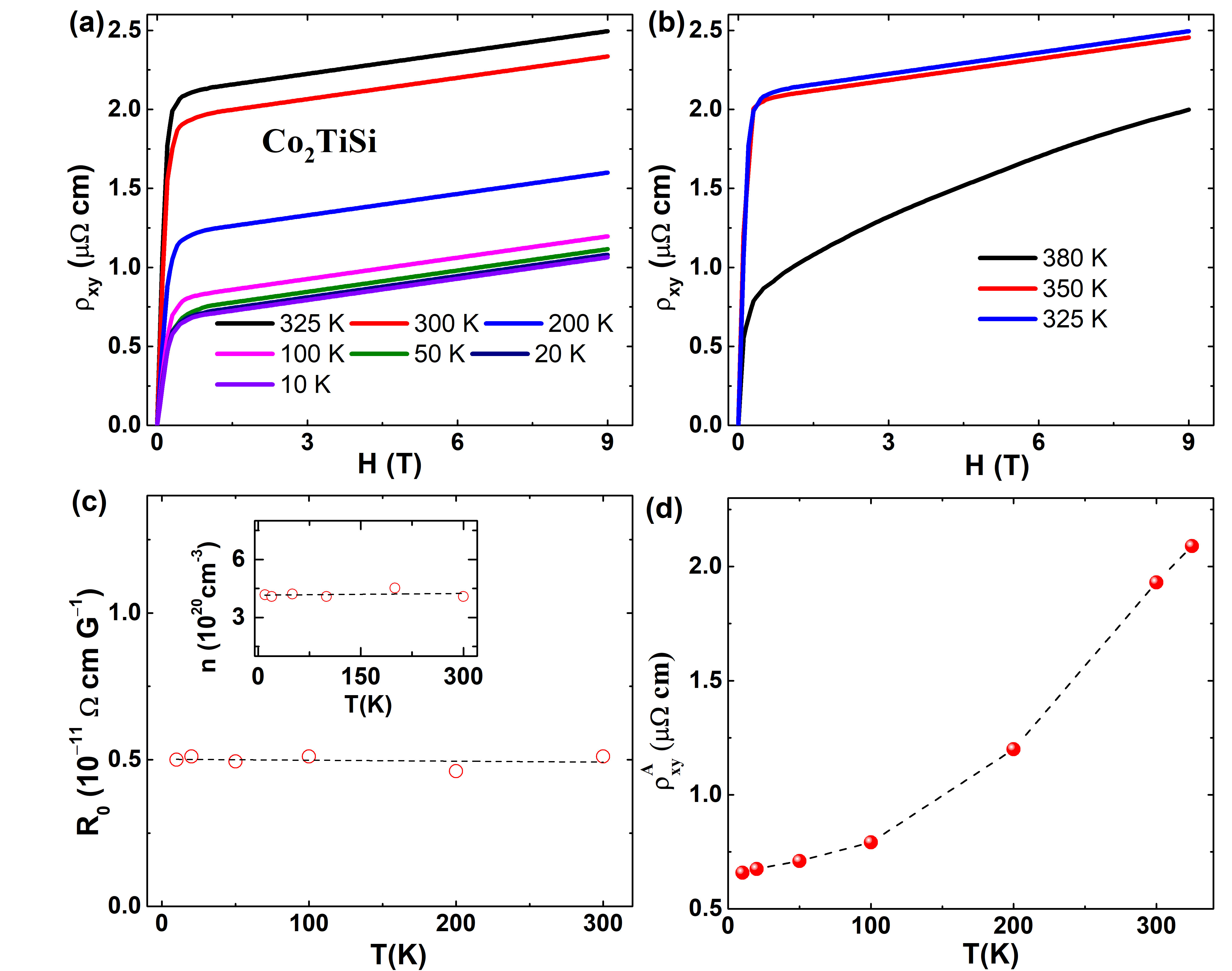}
\caption{(Color online) (a) and (b) Magnetic field dependence of the Hall resistivity ($\rho_{xy}$) for Co$_{2}$TiSi. (c) Temperature dependence of normal Hall coefficient ($R_{0}$). Inset shows the obtained carrier density. (d) Anomalous Hall resistivity ($\rho_{xy}^{A}$) as a function of temperature.}\label{rh}
\end{figure*}

Figures 3(a) and 3(b) illustrate the temperature dependence of longitudinal resistivity for Co$_{2}$TiSi and Co$_{2}$TiGe, respectively. Though the magnetization curves [right insets of Figs. 2(a) and 2(b)] show a clear ferromagnetic to paramagnetic phase transition, we have not observed such behavior in $\rho_{xx}$. Instead, resistivity data for both the compounds exhibit a broad maximum due to the crossover from semiconducting-like to metallic state below the Curie temperature. At 2 K, the values of resistivity are $\sim$109 and $\sim$141 $\mu\Omega$ cm for Co$_{2}$TiSi and Co$_{2}$TiGe, respectively which yield residual resistivity ratio [$\rho_{xx}$(300 K)/$\rho_{xx}$(2 K)] $\sim1.83$ and $\sim1.16$, respectively. In Figs. 4(a) and 4(b), we have plotted the magnetic field dependence of magnetization at various temperatures for Co$_{2}$TiSi and Co$_{2}$TiGe, respectively. Isotherms, well below $T_{C}$, show a sharp increase with increase in field  in the low-field region and a saturation-like behavior starts to appear at high fields. The saturation magnetization for the both materials decreases monotonically with increasing temperature. As we approach towards the Curie temperature, the overall nature of the $M$($H$) curves in the low-field as well as in the high-field regions changes significantly.\\

\begin{figure*}
\includegraphics[width=0.9\textwidth]{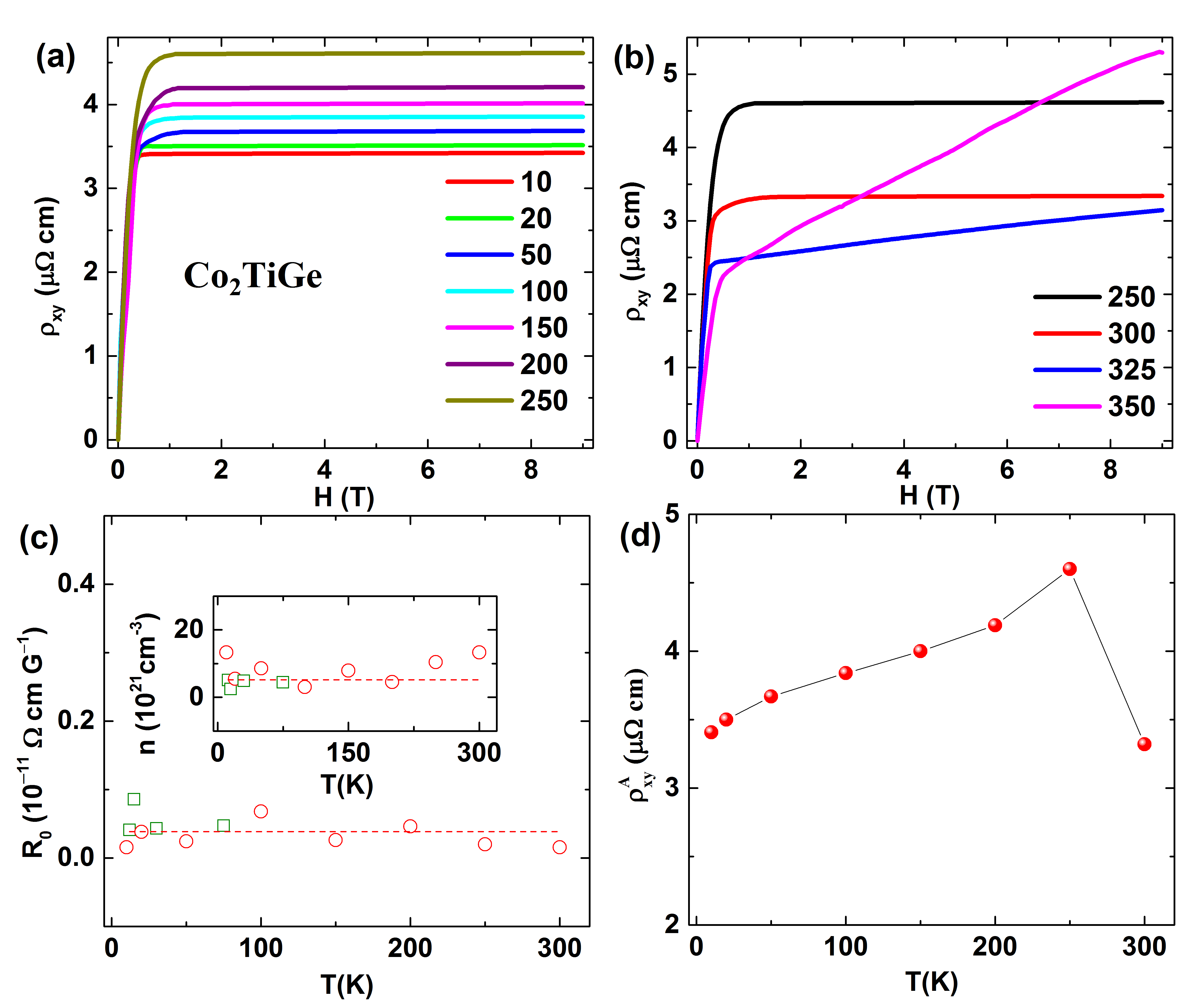}
\caption{(Color online) (a) and (b) Magnetic field dependence of the Hall resistivity ($\rho_{xy}$) for Co$_{2}$TiGe. (c) Temperature dependence of normal Hall coefficient ($R_{0}$). Inset shows the obtained carrier density. (d) Anomalous Hall resistivity ($\rho_{xy}^{A}$) as a function of temperature.}\label{rh}
\end{figure*}

The measured Hall resistivity ($\rho_{xy}$) as a function of magnetic field for different temperatures is shown in Figs. 5(a) and 5(b) for Co$_{2}$TiSi and in Figs. 6(a) and 6(b) for Co$_{2}$TiGe. Similar to $M$($H$), the $\rho_{xy}$($H$) curve has two distinct regions for $T$$<$$T_C$. At low field below $\sim$0.4 T, the Hall resistivity for both the compounds first increases sharply with increase in field. In the high-field region above $\sim$1 T, $\rho_{xy}$ shows a weak linear field dependence up to 9 T. The similarity in the nature of $\rho_{xy}$($H$) and $M$($H$) curves in the low-field region indicates the presence of AHE in these materials. Unlike magnetization, the value of $\rho_{xy}$ increases with the increasing temperature up to $\sim$325 K for Co$_{2}$TiSi and $\sim$250 K for Co$_{2}$TiGe. These temperatures are slightly below the corresponding $T_{C}$s obtained from the magnetisation data. With further increase in temperature, $\rho_{xy}$  decreases gradually.\\

\begin{figure*}
\includegraphics[width=0.9\textwidth]{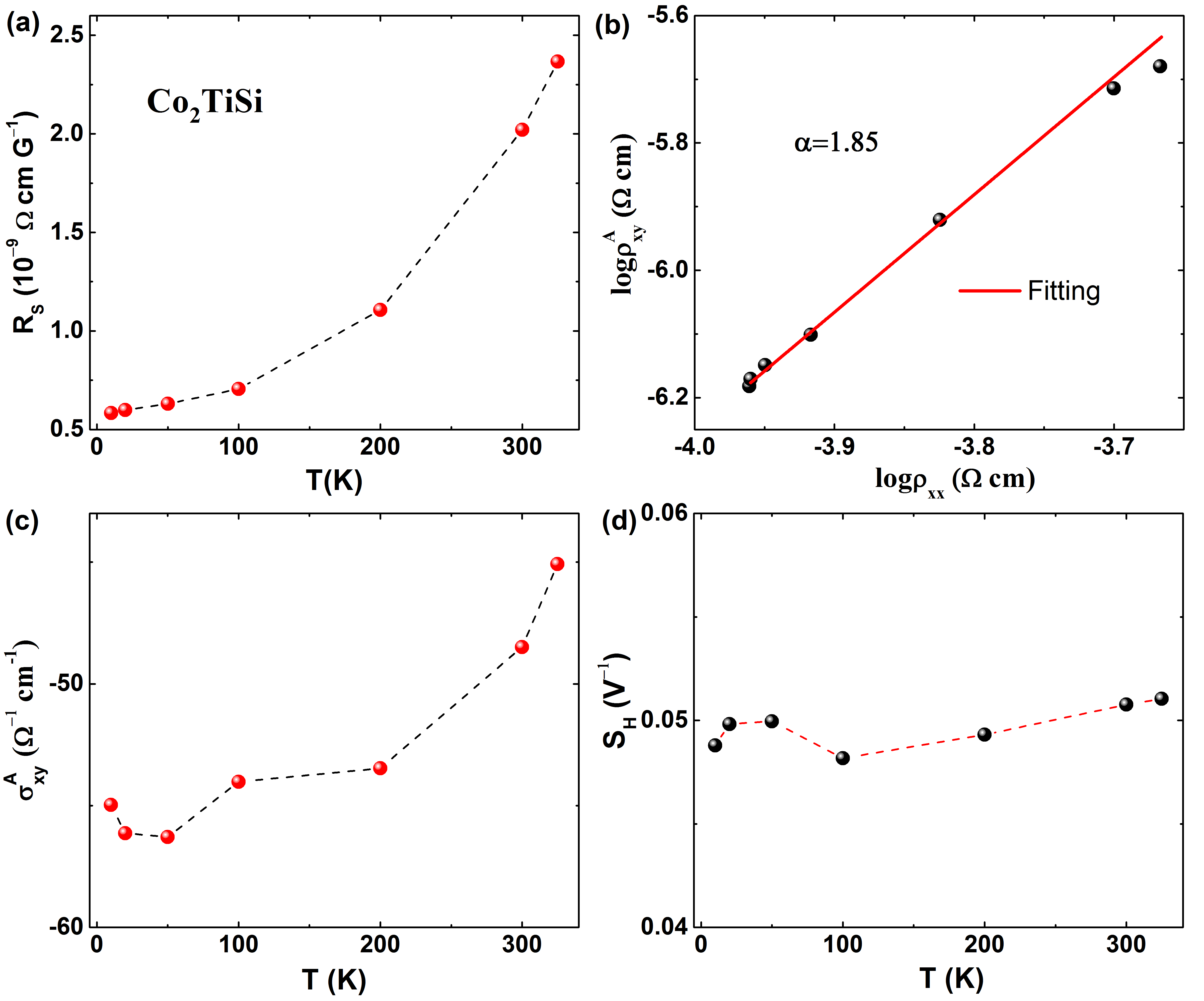}
\caption{(Color online) (a) Temperature dependence of the anomalous Hall coefficient ($R_{S}$) for Co$_{2}$TiSi. (b) Plot of $log$($\rho_{xy}^{A}$) vs $log$($\rho_{xx}$). Solid line is the fitting using the relation, $\rho_{xy}^{A}\propto\rho_{xx}^{\alpha}$. (c) Anomalous Hall conductivity (${\sigma_{xy}}^{A}$) as a function of temperature. (d) Temperature dependence of $S_{H}$.}\label{rh}
\end{figure*}

In addition to usual Hall effect, $\rho_{xy}$ in a ferromagnetic material has a contribution from magnetization and is expressed as,
\begin{equation}
\rho_{xy}=\rho_{xy}^{0}+\rho_{xy}^{A}=R_{0}H+R_{S}\mu_{0}M,
\end{equation}
where $\rho_{xy}^{0}$ is the ordinary Hall resistivity and $R_{0}$ and $R_{S}$ are the ordinary and anomalous Hall coefficients, respectively. Carrier concentration ($n$) can be calculated from $R_{0}=\frac{1}{ne}$, whereas the sign of $R_{0}$ determines the type of the charge carriers. If $M$ saturates, the values of $R_{0}$ and $\rho_{xy}^{A}$ can be derived from the linear fit to $\rho_{xy}$($H$) in the high field region. The slope and $y$-axis intercept of the linear fit then determine the values of $R_{0}$ and $\rho_{xy}^{A}$, respectively. However, it is not possible to perform such analysis, when $M$ does not saturate for magnetic fields below 7 T. So, we can not extract $R_{0}$ and $\rho_{xy}^{A}$ near $T_{c}$. Well below $T_C$, $R_{S}$ can be calculated from the relation, $\rho_{xy}^{A}=R_{S}\mu_{0}M_{S}$. Similar to other ferromagnetic systems, we have extracted $M_{S}$ from the $M$($H$) curves at $H$=5 T \cite{Wang2}. Figures 5(c) and 6(c) show the temperature dependence of $R_{0}$ for Co$_{2}$TiSi and Co$_{2}$TiGe, respectively. Positive values of $R_{0}$ at various temperatures indicate that the transport properties are dominated by hole type charge carriers for both compounds. Using $R_{0}$, we have estimated the carrier concentration [inset of Figs. 5(c) and 6(c) for Co$_{2}$TiSi and Co$_{2}$TiGe, respectively]. It is clear from Figs. 5(c) and 6(c) that the $R_{0}$ values are slightly scattered for both compounds. When the normal Hall coefficient $R_{0}$ is significantly smaller than the anomalous Hall coefficient $R_{S}$, there may be large error in extracted $R_{0}$ and hence in calculated carrier density \cite{Obaida}. For the studied materials, $R_{0}$ is few orders of magnitude smaller than $R_{S}$. At low temperature, $R_{0}$ for Co$_{2}$TiSi is about two orders of magnitude smaller than $R_{S}$. On the other hand, $R_{0}$ for Co$_{2}$TiGe itself is very small and about four orders of magnitude smaller than $R_{S}$. Thus a small error in $R_{0}$ for Co$_{2}$TiGe may affect the calculated value of carrier density significantly. For this reason, we have performed Hall measurements at few more temperatures below 75 K for Co$_{2}$TiGe as compared to the silicon counterpart. \textit{These additional data points are shown using green square-shaped symbol in Fig. 6(c)}. The estimated carrier density for Co$_{2}$TiSi and Co$_{2}$TiGe are $\sim$5$\times10^{20}$ and $\sim$6$\times10^{21}$ cm$^{-3}$, respectively. Thus our Hall measurements confirm that Co$_{2}$TiGe has higher carrier density. From the band structure calculations, Barth \textit{et al.} \cite{Barth} have shown that the total density of states near Fermi energy is almost identical for Co$_{2}$TiSi, Co$_{2}$TiGe, and Co$_{2}$TiSn compounds. Hence, one should expect similar values for the charge carrier density in these materials. We note that the above calculation has been performed without incorporating the SOC. In contrast, recent works suggest that SOC is an essential ingredient for the precise band structure calculation in Co$_{2}$Ti$X$ \cite{Chang,Wang}. The inclusion of SOC leads to Weyl nodes in the momentum space. In fact, other properties of these Weyl nodes are predicted to be tunable with magnetization direction. As the strength of SOC increases with increase in atomic number ($Z$) from Si ($Z$=14) to Ge ($Z$=32) to Sn ($Z$=50), one may find differences in calculated density of states for these compounds, albeit small, when SOC is included. In this context, we would like to mention a recent work on Co$_{2}$TiSn \cite{Ernst}. The carrier concentration for Co$_{2}$TiSn has been reported as $\sim$1.4$\times$10$^{22}$ cm$^{-3}$, which is higher than that for Co$_{2}$TiGe. Thus the carrier concentration increase sequentially with increase of Z from Si to Ge to Sn in this family of materials.\\

\begin{figure*}
\includegraphics[width=0.9\textwidth]{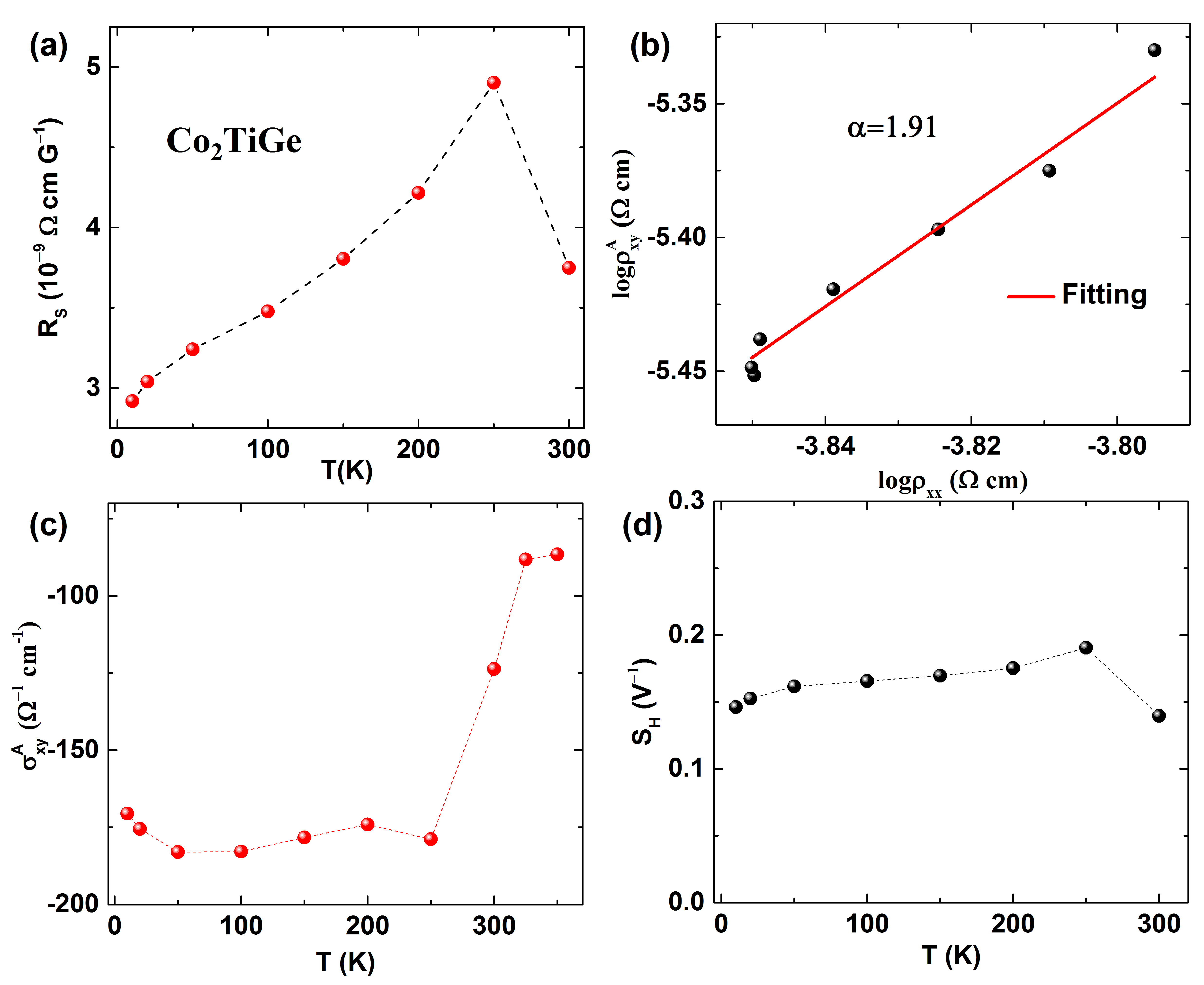}
\caption{(Color online) (a) Temperature dependence of the anomalous Hall coefficient ($R_{S}$) for Co$_{2}$TiGe. (b) Plot of $log$($\rho_{xy}^{A}$) vs $log$($\rho_{xx}$). Solid line is the fitting using the relation, $\rho_{xy}^{A}\propto\rho_{xx}^{\alpha}$. (c) Anomalous Hall conductivity (${\sigma_{xy}}^{A}$) as a function of temperature. (d) Temperature dependence of $S_{H}$.}\label{rh}
\end{figure*}

From Figs. 5(d) and 6(d), it is evident that the anomalous component of the Hall resistivity increases monotonically with increase in temperature from 10 K  up to $\sim$325 K for Co$_{2}$TiSi and up to $\sim$250 K for Co$_{2}$TiGe. In Co$_{2}$TiGe, $\rho_{xy}^{A}$ decreases above 250 K. This is also reflected in the calculated anomalous Hall coefficient $R_{S}$ for both materials [Figs. 7(a) and 8(a)]. In Figs. 7(b) and 8(b), we have checked the scaling behavior of $\rho_{xy}^{A}$ using the relation, $\rho_{xy}^{A}=\beta\rho_{xx}^{\alpha}$, in the temperature range 10 to 325 K and 10 to 250 K for Co$_{2}$TiSi and Co$_{2}$TiGe, respectively. For both the systems, almost quadratic dependence of $\rho_{xy}^{A}$ on $\rho_{xx}$ ($\alpha$$\approx$2) confirms that the AHE is originated from either intrinsic KL or extrinsic side jump mechanism.\\

We have also calculated anomalous Hall conductivity for understanding its possible origin. Figures 7(c) and 8(c) show the temperature dependence of corresponding anomalous Hall conductance (AHC) $(\sigma_{xy}^{A}\approx-\frac{\rho_{xy}^{A}}{\rho_{xx}^{2}}\approx-\frac{R_{S}\mu_{0}M_{S}}{\rho_{xx}^{2}})$ for Co$_{2}$TiSi and Co$_{2}$TiGe, respectively. At 20 K, $\mid\sigma_{xy}^{A}\mid$ is about 55 $\Omega^{-1}$ cm$^{-1}$ for Co$_{2}$TiSi and $\sim$170 $\Omega^{-1}$ cm$^{-1}$ for Co$_{2}$TiGe. With increasing temperature, $\mid\sigma_{xy}^{A}\mid$ gradually decreases. Ideally, when a nodal-line in the electronic band structure is almost dispersion-less and Fermi level resides in the band gap induced by SOC, i.e., in the resonance condition, the intrinsic contribution to $\sigma_{xy}^{A}$ resonantly enhances and approaches towards a value $\frac{e^{2}}{ha}$ \cite{Onoda,Miyasato,KKim}, where $e$, $h$, and $a$ are the electronic charge, Planck constant, and lattice constant, respectively. Using $a$=5.811 ${\AA}$, we have estimated the intrinsic contribution to be $\mid\sigma_{xy,in}^{A}\mid$=666 $\Omega^{-1}$ cm$^{-1}$ for Co$_2$TiGe, which is comparable to the experimentally obtained value of $\sigma_{xy}^{A}$ ($\sim$170 $\Omega^{-1}$ cm$^{-1}$). Whereas in the case of Co$_2$TiSi, $\sigma_{xy}^{A}$ ($\sim$ 55 $\Omega^{-1}$ cm$^{-1}$) is one order of magnitude smaller than the calculated $\mid\sigma_{xy,in}^{A}\mid$=670 $\Omega^{-1}$ cm$^{-1}$. In this context, it may be mentioned that anomalous Hall effect has also been studied in Co$_2$TiSn compound. For Co$_2$TiSn, the deduced value of $\sigma_{xy}^{A}$ is 284 $\Omega^{-1}$ cm$^{-1}$ at low temperature region \cite{Ernst}. This value of $\sigma_{xy}^{A}$ is larger than that for Co$_2$TiGe. Thus both carrier density and anomalous Hall conductivity in Co$_2$Ti$X$ series increase systematically from Si to Sn. As one might anticipate, the nodal-lines are not always dispersive along the expected direction, and only the part of the nodal-line, which satisfies the resonance condition, contributes to the intrinsic anomalous Hall conductivity. As a result, the theoretically calculated $\mid\sigma_{xy,in}^{A}\mid$ generally deviates from the idealized value $\frac{e^{2}}{ha}$. There are several examples \cite{KKim,Wang3,Liu}, where the experimental values of $\mid\sigma_{xy}^{A}\mid$ show significant deviation from the theoretically calculated $\mid\sigma_{xy,in}^{A}\mid$, in spite of being originated from intrinsic Berry phase. On the other hand, the extrinsic side jump contribution to $\sigma_{xy}^{A}$ should be of the order of $\frac{e^{2}}{ha}(\frac{\varepsilon_{SO}}{E_{F}})$, where $\varepsilon_{SO}$ and $E_{F}$ are SOC and Fermi energy, respectively. We note that $\frac{\varepsilon_{SO}}{E_{F}}$ is usually less than 0.01 for ferromagnetic metals \cite{Wang2,KKim}. Hence, the intrinsic Berry phase driven KL contribution should dominate the AHC in Co$_2$Ti$X$. Nonetheless, there can be an extrinsic side-jump contribution, albeit small. However, in practice, it is not possible to decouple these two components.\\

As intrinsic AHC, $\mid\sigma_{xy,in}^{A}\mid$, is approximately proportional to magnetization, the scaling coefficient, $S_{H}=\frac{R_{S}\mu_{0}}{\rho_{xx}^{2}}$=$\frac{\sigma_{xy,in}^{A}}{M}$ should be constant and independent of temperature \cite{Wang2,Zeng}. Figures 7(d) and 8(d), for Co$_{2}$TiSi and Co$_{2}$TiGe, respectively, show that $S_{H}$ is almost constant with temperature, which also confirms the dominant intrinsic Berry phase contribution to AHC. The obtained value of $S_{H}$ is comparable to that reported previously for several Heusler compounds and ferromagnetic materials \cite{Wang2,Husmann}. The half-metallic magnetic Heuslers Co$_2$Ti$X$ are predicted to host Weyl fermions associated with the large Berry phase of their Fermi surfaces \cite{Chang,Wang}. The enhanced Berry curvature near the band crossings at $E_{F}$ is expected to add a large contribution to the intrinsic AHE. Though several theoretical works have been performed in order to correlate AHC with the Berry phase curvature \cite{Nagaosa2,Jungwirth,Kubler,Yao}, very few experimental observations have been reported on the AHE of Heusler alloys to confirm the theoretical prediction \cite{Husmann,Ludbrook}. Recently, Berry curvature calculations have been performed for Co$_{2}$TiSn near the Fermi energy and intrinsic anomalous Hall conductivity of 100 $\Omega^{-1}$ cm$^{-1}$ has been calculated \cite{Ernst}. The anomalous Hall conductivity in Co$_{2}$TiSn mainly comes from slightly gaped nodal lines by magnetization-induced symmetry reduction. However, the experimentally obtained AHC in Co$_{2}$TiSn thin film was attributed to intrinsic KL mechanism associated with the Berry phase as well as the extrinsic mechanisms. All the same, recently, some ferromagnetic and non-collinear antiferromagnetic metals have been confirmed to show the intrinsic Berry phase driven AHC \cite{Wang2,Nayak,Nakatsuji,Kiyohara}. The detail analysis of AHE of the half-metallic full Heusler Co$_2$Ti$X$ shows that $\rho_{xy}^{A}$ is primarily originated from the intrinsic KL mechanism associated with the Berry phase. The experimentally obtained AHC is comparable to that reported for several ferromagnetic and non-collinear antiferromagnetic materials \cite{Nayak,Nakatsuji,Kiyohara}.

\section{Conclusion}

In summary, we have performed transport and magnetic measurements on Co$_{2}$Ti$X$. These half-metallic Heusler compounds are predicted to host Weyl fermions with broken time reversal symmetry along with the large associated Berry phase of their Fermi surfaces. In Co$_{2}$Ti$X$, the enhanced Berry curvature near the $E_{F}$ is expected to add a large contribution to the intrinsic AHE. We have carried out a detailed analysis of AHE for Co$_{2}$TiSi and Co$_{2}$TiGe compounds in which a scaling relation between $\rho_{xy}^{A}$ and $\rho_{xx}$ has been obtained as $\rho_{xy}^{A}\propto \rho_{xx}^\alpha$, $\alpha\sim2$. The almost quadratic dependence ($\alpha$$\approx$2) suggests that AHE is originated from either intrinsic KL or extrinsic side jump mechanism.  The experimentally obtained AHC found to be comparable with the theoretically calculated intrinsic AHC in Co$_{2}$TiGe, whereas, it is one order smaller in Co$_{2}$TiSi. However, $S_{H}$ is almost temperature independent for both the materials, which supports intrinsic nature of AHC. The large AHC and AHE make Co$_{2}$Ti$X$  promising candidates for future spintronics applications. They also provide a platform for further investigations of exotic properties in other Co$_{2}$-based Heusler materials.

\section{Acknowledgments}

We thank Mr. Arun Kumar Paul for his help during sample preparation and measurements.

\end{document}